\documentclass[secnumarabic,aps,prb,reprint]{revtex4-2}
\usepackage[english]{babel}
\usepackage{graphicx}
\usepackage{color,soul}
\usepackage{amsmath}
\usepackage{tabularx}
\usepackage[dvipsnames]{xcolor}

\begin{document}

	\title{Atomic and inter-atomic orbital magnetization induced in SrTiO$_3$ by chiral phonons}
	\author{Sergei Urazhdin}
	\affiliation{Department of Physics, Emory University, Atlanta, GA, USA}

\begin{abstract}
An unexpectedly large transient magnetization induced by circularly polarized ferroelectric phonons was recently observed in a nonmagnetic insulator SrTiO$_3$ [Nature \textbf{628}, 534 (2024)]. We use a minimal molecular orbital model to demonstrate two electronic contributions to this effect. An atomic orbital contribution arises from the pumping of orbital angular momentum of Ti by chiral motion of coordinating oxygen atoms. An additional inter-atomic contribution is associated with the transient circulating current around the oxygen atoms, resulting in efficient dressing of phonons by electron dynamics. The insights provided by our model may facilitate the development of ultrafast magnetization control and orbitronic sources.
\end{abstract}

\maketitle

\begin{figure}
	\centering
	\includegraphics[width=\columnwidth]{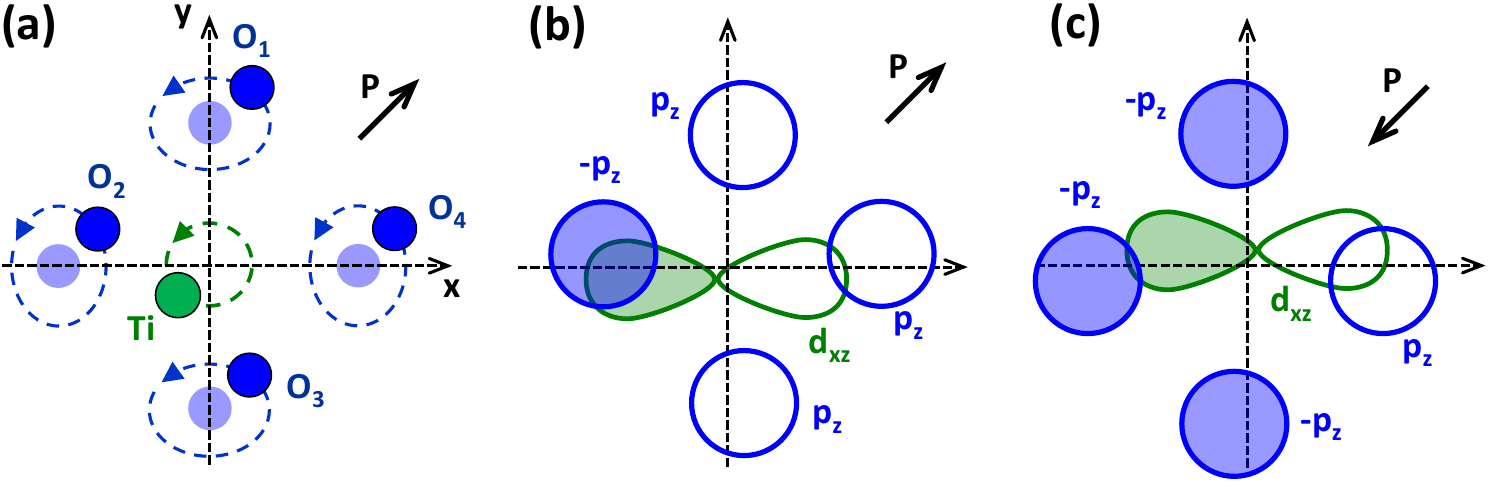}
	\caption{\label{fig:STO_phonon} (a) Schematic of atomic trajectories in the circularly polarized ferroelectric phonon mode. (b),(c) bonding molecular orbitals derived from the Ti $d_{xy}$ orbital for two opposite phases of atomic motion in the adiabatic approximation.}
\end{figure}

\section{Introduction} 

Basini {\textit et al}~\cite{Basini2024} recently showed that chiral ferroelectric phonons generated in strontium titanate SrTiO$_3$ (STO) by circularly polarized THz pulses produce large transient magnetization detected as a magneto-optic Kerr effect (MOKE) signal. A DC signal was observed in addition to an oscillating signal due to the THz ionic Kerr effect~\cite {PhysRevB.109.024309}. It was interpreted as the dynamical Barnett effect $-$ the inverse of the Einstein-de Haas effect $-$ in which mechanical rotation of a solid induces magnetization $M_{BE}$~\cite{Barnett1948}. In the simplest interpretation, opposite-phase circulating motion of ions produces circulating local currents and thus magnetic moments [Fig.~\ref{fig:STO_phonon}(a)]. However, the estimated ionic effect was too small, suggesting that circularly polarized phonons may efficiently couple to the electronic subsystem inducing spin and/or orbital electron polarization. 

In a related experiment, magnetization of a transition metal-rare earth alloy ferrimagnet (FIM) film was reversed via optical excitation of circularly-polarized phonons in Al$_2$O$_3$ and SiO$_2$ substrates~\cite{Davies2024}. This effect was attributed to either the Ampere field driven by the dynamical lattice polarization, or alternatively to the electronic magnetization induced by transfer of angular momentum from the lattice via the dynamical crystal field or spin-orbit coupling (SOC). The lack of the accepted mechanism for these effects reflects the complexity of coupled electron-phonon dynamics, which is outlined below as the motivation for the proposed model.

\textbf{Ionic magnetic moment.} The most tractable of the discussed mechanisms is the Ampere field produced by the rotating ion polarization $\mathbf{P}$~\cite{PhysRevMaterials.1.014401,Rckriegel2020}. $\partial_t\mathbf{P}$ gives the current density, so the magnetization is proportional to $\mathbf{P}\times\partial_t\mathbf{P}$. The relation between the magnetization and angular momentum density carried by the  phonon is
$\boldsymbol{\mu}_{uc}=\sum_i\gamma_i\mathbf{L}_i$,
where $L_i=M_i\mathbf{Q}_i\times\dot{\mathbf{Q}}_i$ is the angular momentum of ion $i$ in a unit cell (u.c.), $\gamma_i=\sum_i eZ_i/2M_i$ is its gyromagnetic ratio, $Z_i$ is the Born effective charge of ion $i$, $M_i$ is mass, and $\mathbf{Q}_i$ is displacement~\cite{PhysRevMaterials.1.014401,PhysRevMaterials.3.064405}. 

To evaluate this effect for Al$_2$O$_3$, we assume that $Z_{Al}$, $Z_{O}$ are isotropic and the same for all 12 Al and 18 O atoms in the u.c. The magnetic moment is then $\mu_{uc}=\omega/2\sum_ieZ_iQ_i^2$, where $\omega$ is frequency. For $Al_2O_3$, $\omega\sim10^{14}$~s$^{-1}$ (wavelength $20\,\mu$m), $Z_{Al}\approx-3$, $Z_{O}\approx 2$, and $Q_{O}=-3M_{Al}Q_{Al}/2M_{0}\approx-1.1Q_{Al}$. Evaluation of displacement amplitude for the large optical field $E>10^5$~V/cm$^2$ used in Ref.~\cite{Davies2024} requires nonlinear analysis of lattice dynamics. For an estimate, we use $Q_{Al}=0.1\,\AA$ corresponding to the relative Al-O displacement amplitude of $10\%$. The resulting magnetic moment is $\mu_{uc}\approx5\times10^{-27}$~J/T ($5\times10^{-4}\mu_B$), and the magnetization is $M=\mu_{uc}/V\sim20$~A/m, where $V=254\,\AA^{3}$ is u.c. volume. This results in a field $B=\mu_0M=2\times10^{-5}$~T inside the magnetized substrate volume. At the location of the FIM film above this volume, the field is produced only by the fringe effects determined by the spatial profile of magnetization, which is smaller. We arrive at the same conclusion as Ref.~\cite{Basini2024}, that ionic effects alone may be insufficient to explain the observations.

Analysis of angular momentum transfer from the lattice to electrons is complicated by the distinction between the true angular momentum $\mathbf{J}$ of phonons, which is conserved for an isolated solid, and their angular quasi-momentum $\mathbf{J}_Q$~\cite{Rckriegel2020,PhysRevB.103.L100409}. The latter arises in the analysis of quasiparticle scattering on the lattice~\cite{PhysRevB.97.174403}, but similarly to the Bloch momentum is generally not conserved. This quantity also does not account for rotations of the entire lattice, so it cannot describe Einstein-de Haas or Barnett effects. In the studies of molecular dynamics, this problem is addressed by separating rigid rotations and vibrations, known as Eckart convention~\cite{RevModPhys.69.213}. 

\textbf{Electronic magnetism driven by lattice dynamics.} Angular momentum of Bloch electrons on the static lattice is well understood, although the significance of orbital contribution has been recognized only recently~\cite{Go2021-sn}. However, its coupling to angular momentum of lattice dynamics can be complex. Nonlinear multi-phonon processes are essential for angular momentum conservation in spin-lattice interaction~\cite{Garanin2021}. Furthermore, orbital moment of electrons can relax into the lattice on femtosecond timescale~\cite{PhysRevB.108.L180404}, which can lead to a non-perturbative coupling regime where angular momentum is collectively carried by the electron-phonon coupled system and is described by a new quasiparticle, the angulon~\cite{Mentink2019}.

Spin-lattice interaction has been extensively studied in the context of spin relaxation~\cite{1961}. In Van Vleck's mechanism, spin interacts with the lattice via magnetic dipole field of phonons~\cite{PhysRev.57.426}.  Alternatively, interaction can be mediated by SOC. For magnetic solids, this can result in direct transfer of angular momentum between phonons and electron spins, which is facilitated by the almost gapless spectrum of spin excitations~\cite{Rckriegel2020}. First principles calculations suggest that spin-lattice interaction can be dominated by the dynamical lattice inversion symmetry breaking, which in magnetic materials results in Dzyaloshinskii-Moriya-type spin interactions~\cite{Mankovsky2022}. These mechanisms are relevant to paramagnetic or ferromagnetic spins, but not non-magnetic STO.

Coupling of electron orbital magnetization to lattice dynamics does not require SOC. The geometric mechanism is associated with the Berry phase of the wavefunction cyclically perturbed by the lattice dynamics which can produce magnetization even in insulators~\cite{Thonhauser2005,PhysRevB.100.054408,Zhang2023}, making this mechanism potentially relevant to the discussed effects. Orbital magnetization is expected to scale linearly with the displacement amplitude, which appears to contradict the quadratic dependence of magnetization on field amplitude observed in Ref.~\cite{Basini2024}. 

Chiral phonon-induced electron magnetism observed in paramagnetic rare earth compounds~\cite{Luo2023} was explained by the splitting of the 4f levels by the dynamical crystal field, which plays the role of a large effective magnetic field efficiently polarizing a partially occupied SOC electronic level manifold~\cite{Juraschek2022}. However, this mechanism is not directly applicable to insulators such as STO with quenched spin and orbital degrees of freedom. The atomic shells of all three atomic species in STO are filled or empty, so their Zeeman-like splitting by the dynamical crystal field does not in itself result in spin or orbital polarization. In the electronic band picture all the bands are either filled or empty, and are consequently magnetically and electrically inert. A large band gap of $\Delta=3.2$~eV between the filled valence and the empty conduction bands prevents phonon-driven electronic excitation. The energy of the ferroelectric phonon  with frequency $f=\omega/2\pi=3$~THz is only  $E_{ph}=\hbar\omega=12.4$~meV. Linear excitation via single-phonon absorption is not possible, as it requires $E_{ph}>\Delta$. Because of the vast difference of energy scales, perturbative nonlinear (multi-phonon) effects also cannot explain the observations.

Given the complexity and diversity of possible interpretations of phenomena involving angular momentum of coupled electron-phonon system, one may ask whether it is possible to develop a minimal model that allows one to identify the dominant effects. In this paper, we address this question by considering a simple molecular orbital approximation for the valence states in STO. Our minimal model shows that chiral phonons produce both intra- and inter-atomic orbital polarization of electrons due to transient dynamical reconstruction of electronic states, renormalizing the magnetic moments of phonons. The microscopic insights provided by our model may facilitate the development of methods to efficiently control magnetization and generate orbital moments in orbitronic applications via lattice dynamics~\cite{PhysRevLett.102.016601}.

\section{Analysis and Results} 

\subsection{Molecular orbital model}

Atomic motion in the ferroelectric mode impacts predominantly the valence electronic states forming conduction and valence bands, because both the energies and the populations of the latter are sensitive to atomic displacements. In STO, these states are derived from the 2p orbitals of oxygen and $t_{2g}$ 3d-orbitals of Ti, with negligible contribution from Sr~\cite{PhysRev.135.A1321,PhysRevB.6.4718}. One can thus focus on the effects of gyration of oxygen atoms relative to Ti they coordinate. We assume that the driving optical field is incident in the z-direction, resulting in gyration in the xy plane, Fig.~\ref{fig:STO_phonon}(a). Since the dynamics is quasi-uniform through the crystal, it does not produce charge currents across unit cells. Therefore, molecular orbital (MO) approximation can provide a minimal model for the dynamical effects. 

The significance of coupling between atomic motion and the electronic state in STO becomes  apparent by considering, for example, the bonding MO derived from the Ti $d_{xz}$ orbital. 
We assume that in equilibrium, all the Ti-O bonds are equal, as expected for the room-temperature cubic phase. Without ferroelectric distortion, the $d_{xz}$ orbital of Ti hybridizes only with the $p_z$ orbitals of the two oxygen atoms on the x-axis~\cite{Dylla2019}. In the presence of ferroelectric distortion, it becomes hybridized with the $p_z$ orbitals of the two oxygen atoms on the y-axis, with the orbital phases dependent on the instantaneous direction of displacement vector $\mathbf{Q}=Q(\cos\omega t,c\sin\omega t)$, where $c=\pm 1$ is the mode chirality, Figs.~\ref{fig:STO_phonon}(b),(c). The same picture holds for the MO derived from the d$_{yz}$ orbital, with the x- and y-axes exchanged, i.e., strong hybridization along the y-axis, and alternating hybridization along the x-axis. For superpositions of $d_{xz}$ and $d_{yz}$ orbitals, dynamical hybridization along both axes results in the oscillating relative phase, generating orbital states $d_\sigma=-(\sigma d_{xz}+id_{yz})/\sqrt{2}$ characterized by unquenched orbital moment $\sigma=\pm1$ along the z-axis.

The SOC neglected in the above analysis mixes the d$_{xz}$ and d$_{yz}$ orbitals into $d_\sigma$ and their superpositions with $d_{xy}$ orbitals entangled with spin, forming Kramers doublets~\cite{Urazhdin_STO}. For instance, one such doublet is $d_{+}|\downarrow\rangle$, $d_{-}|\uparrow\rangle$. Thus, it is sufficient to analyze the effects of dynamical hybridization on the MOs derived from the states $d_\sigma$, with the spin playing the role of a passive index determined by the orbital state. The Ti spin-orbit splitting $\lambda=17$~meV is negligible compared to the insulating gap $\Delta=3.2$~eV defining the characteristic energy scale. A finite amplitude of d$_{xy}$ in some of the SOC states does not affect the mechanism discussed below.

\begin{figure}
	\centering
	\includegraphics[width=\columnwidth]{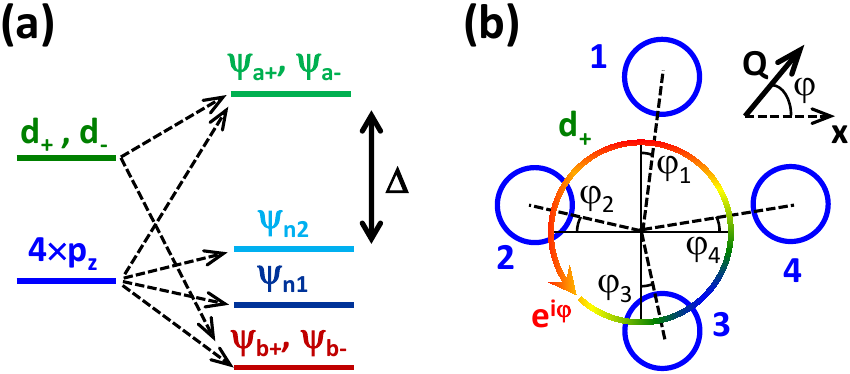}
	\caption{\label{fig:STO_chiral} (a) MO crystal-field splitting of the levels derived from the considered atomic orbitals. $\Delta$ represents the bandgap. (b) Illustration of the mechanism for the complex phase acquired by the matrix elements between Ti $d_\sigma$ and O $p_z$ orbitals due to the ferroelectric displacement.}
\end{figure}

We now consider the MO states formed by Ti $d_\sigma$ orbitals and the $p_z$ orbitals of the four oxygen atoms coordinating Ti in the xy plane. We count the  energy relative to the atomic oxygen p-level. In the absence of ferroelectric distortion, the two-fold degenerate Ti atomic levels $d_\sigma$ with energy $E_d$ and the four degenerate  $p_z$-levels of oxygen are split due to hybridization into two orbitally non-degenerate and two orbitally degenerate states, Fig.~\ref{fig:STO_chiral}(a). The non-degenerate states are the bonding and the antibonding states of the four oxygen atoms, both non-bonding with Ti,
\begin{equation}\label{eq:nb}
	\psi_{n1}=\frac{1}{2}\sum p_{n},\ \psi_{n2}=\frac{1}{2}\sum (-1)^np_{n},
\end{equation}
where $p_{n}$ denotes the $p_z$ orbital of $n^{th}$ oxygen, $n=1..4$ [see Fig.~\ref{fig:STO_phonon}(a)]. Their energies are $E_{n1,n2}=\pm 2t_{O-O}$, where $t_{O-O}<0$ is the matrix element between $p_z$ orbitals of the nearest-neighbor oxygen atoms. The two orbitally degenerate states are the antibonding and the bonding states with respect to the Ti-O hybridization,
\begin{equation}\label{eq:ab}
		\psi_{a\sigma,b\sigma}=\frac{\cos\theta_{a,b}}{2}\sum e^{in\sigma\pi/2}p_n+\sin\theta_{a,b}d_\sigma,
\end{equation}
with energies
\begin{equation}
E_{a,b}=\frac{E_d}{2}\pm\sqrt{\frac{E_d^2}{4}+2t^2_{Ti-O}},
\end{equation}
where $t_{Ti-O}<0$ denotes the hopping matrix element from the $p_z$ orbital of oxygen No.$4$ to $d_\sigma$ orbital of Ti, $t_{Ti-O}=\langle d_\sigma|\hat{H}|p_4\rangle$, where $\hat{H}$ is the Hamiltonian. In Eq.~(\ref{eq:ab}), we use trigonometric parametrization with $\theta_{a,b}=\tan^{-1}\frac{\sqrt{2}t_{Ti-O}}{E_d-E_{a,b}}$ to simplify state normalization. The Ti-O bonding states and the non-bonding states are filled, while the antibonding state is empty, with the gap 
\begin{equation}
\Delta=\frac{E_d}{2}+\sqrt{\frac{E_d^2}{4}+2t^2_{Ti-O}}+2t_{O-O}
\end{equation}
between the antibonding and the nonbonding state $\psi_{n2}$ representing the bandgap of STO, Fig.~\ref{fig:STO_chiral}(a).

We now analyze the effect of ferroelectric displacements on the hybridization and the resulting perturbation of the MO states. Rotation of the displacement vector results in the periodic modulation of both the bond length and its direction. The former entails modulation of the matrix element amplitude, while the latter $-$ complex phase due to the chirality of the $d_\sigma$ orbitals, Fig.~\ref{fig:STO_chiral}(b). Taking into account the directional dependence of the matrix element phases in the symmetric state, the perturbation of the matrix elements describing hopping from the Ti orbital $d_{\sigma}$ to $n^{th}$ oxygen is
\begin{equation}\label{eq:dt1}
	\delta t_{n,\sigma}=Qe^{i\sigma n\pi/2}[a_l\cos\xi+i\sigma a_t\sin\xi],
\end{equation}
where $\xi=\varphi-n\pi/2$, $\varphi$ is the polar angle of instantaneous displacement and $a_l,a_t$ are real parameters with units of force describing the effects of displacement $\mathbf{Q}$ in the direction of and normal to the bond, respectively.  Since $t_{Ti-O}<0$ and $Q>0$, $\xi=0$ describes increased length of the Ti-O$_4$ bond, so $a_l>0$. Likewise,  $Q>0$, $\xi=\pi/2$ describes a shift of O$_4$ in the $+y$ direction, which is equivalent to the  positive (negative) phase shift of the orbital $d_+$ ($d_-$), corresponding to $a_t>0$ in Eq.~(\ref{eq:dt1}). These parameters can be obtained from \textit{ab-initio} calculations or by the Koster-Slater method, as discussed below.

For the circularly polarized phonon mode, $\varphi=c\omega t$ with chirality $c=\pm 1$. Then
\begin{equation}\label{eq:dt2}
	\delta t_{n,\sigma}=Qa_+e^{ic\sigma\omega t}+(-1)^nQa_-e^{-ic\sigma\omega t}.
\end{equation}
where we use $a_\pm=(a_l\pm a_t)/2$. The first term is isotropic, enabling coupling between the isotropic non-bonding state $\psi_{n1}$ [Eq.~(\ref{eq:nb})] and $\psi_{a,\sigma}$. The corresponding matrix element is $V_{\sigma,n1}=2\sin\theta_aQa_+e^{-ic\sigma\omega t}$. Meanwhile, the second term in Eq.~(\ref{eq:dt2}) couples $\psi_{a,\sigma}$ to $\psi_{n2}$, with the matrix element $V_{\sigma,n2}=2\sin\theta_aQa_-e^{ic\sigma\omega t}$.

\subsection{Transient Atomic Orbital Magnetism}

We now show that time-dependent hybridization associated with the chiral phonon results in transient atomic polarization. If $a_+$ and $a_-$ were equal, then $V_{\sigma,n2}=V_{-\sigma,n1}$ and phonons would not break the symmetry between the orbitals $d_+$ and $d_-$ carrying orbital moment. However, since both $a_l$ and $a_t$ are positive, $a_+>a_-$. The first term in Eq.~(\ref{eq:dt2}) is dominant, resulting in a chiral perturbation due to different phonon-driven admixing of $\psi_+$ and $\psi_-$ to the non-bonding states. Qualitatively, the factor $e^{\pm i\omega t}$ either increases or decreases the time-dependent phase difference between $\psi_{n1,n2}$ and $\psi_{a,\sigma}$ orbitals, resulting in the dependence of MO mixing on the product of phonon chirality $c$ and orbital moment $\sigma$.

We use time-dependent perturbation theory~\cite{Landau1981-ae} with respect to $\delta t_{n,\sigma}$ to analyze this effect. Approximating the time dependence of phonon amplitude by $Q(t)=Qe^{-|t|/\tau}$ with the temporal pulse width $\tau\gg\hbar/\Delta$, the probability to find an electron in the state $\psi_{a,\sigma}$ at the polarization maximum at $t=0$ is

\begin{equation}\label{eq:p_sigma_0}
	p_\sigma(t=0)=-\frac{4\sin^2\theta_aQ^2_0(a^2_+-a^2_-)}{(\Delta-c\sigma\hbar\omega)^2+\hbar^2/\tau^2},
\end{equation}
where we have neglected the contribution of $t_{O-O}$ to the denominator dominated by the Ti-O bonding. Accounting for the two possible directions of spin doubling the considered orbital effect and using $1/\tau\ll\Delta/\hbar$, the induced orbital moment $m_z(t=0)=-2\mu_B(p_+-p_-)$ is
\begin{equation}\label{eq:moment}
	m_{z,a}(t=0)\approx\frac{32\mu_Bc\hbar\omega\sin^2\theta_aQ^2_0(a^2_+-a^2_-)}{\Delta^3},
\end{equation}
where $\mu_B$ is Bohr magneton. The magnetic moment is quadratic in the displacement amplitude $Q$, consistent with the observations of Ref.~\cite{Basini2024}. Note that $(a^2_+-a^2_-)=a_la_t$, showing that an unquenched orbital state required for a finite $a_t$ is essential for electron magnetization.

The experiment of Ref.~\cite{Basini2024} also showed that the magnitude of induced magnetization closely followed the phonon amplitude, disappearing together with the latter. Our analysis reproduces this transient behavior as well. According to the time-dependent perturbation theory, the probability that electron is found in the state $\psi_{a,\sigma}$ at  $t\gg\tau$ is
\begin{equation}\label{eq:p_sigma_inf}
	p_\sigma(t=\infty)=\frac{16\hbar^2\sin^2\theta_aQ^2_0(a^2_+-a^2_-)}{\tau^2(\Delta-c\sigma\hbar\omega)^4}.
\end{equation}
Using the experimental parameters of Ref.~\cite{Basini2024}, this value is scaled by the factor $4\tau^2\Delta^2/\hbar^2\sim6\times10^{-5}$ compared to the maximum transient amplitude Eq.~(\ref{eq:p_sigma_0}). In other words, the probability that an electron is actually excited from the valence band into the conduction band due to the interaction with phonons is negligible. Transient magnetization cannot be described as electronic excitation driven by the electron-phonon interaction in the rigid-band approximation. Instead, it can be interpreted as a virtual excitation of conduction-band electrons, i.e. a transient state that results from the dynamical evolution of the band states coupled non-perturbatively to the lattice dynamics, which vanishes together with the phonon-driven mixing of electronic states. Analysis of electron-phonon coupling in several other materials arrived at a similar conclusion~\cite{PhysRevB.110.094401,PhysRevB.107.L020406,Zhang2023}, suggesting that it may be quite general.

We now estimate the relationship between induced magnetization and the amplitude $Q$ of atomic displacement. The antibonding mode $\psi_{a,\sigma}$ describing the conduction band states is dominated by the Ti $d$-orbitals, i.e. $\sin\theta_a\lesssim1$. We evaluate the effective forces $a_+$, $a_-$ using the Koster-Slater method~\cite{Harrison}. The value of $a_t$ is determined by the bonding energy geometrically scaled by the directional dependence of orbital phase, 
\begin{equation}
a_t=-t_{Ti-O}/r_{Ti-O}\approx5.7\,eV/nm,
\end{equation}
where $r_{Ti-O}=0.2$~nm is the titanium-oxygen distance, and $t_{Ti-O}=-V_{pd\pi}=1.14$~eV with the standard notation for the hopping matrix element between the $d$ and $p$ orbitals~\cite{Harrison}. Meanwhile, $a_l$ is the derivative of the bond energy with respect to the oxygen-titanium distance,
\begin{equation}
a_l=\frac{dt_{Ti-O}(r_{Ti-O})}{dr_{Ti-O}}=\frac{7a_t}{2}=20.0\,eV/nm,
\end{equation}
where we used the semi-empirical dependence $t_{Ti-O}(r_{Ti-O})\propto r_{Ti-O}^{-7/2} $~\cite{Harrison}. Thus, we estimate $a_+\approx 13$~eV/nm, $a_-\approx 7$~eV/nm. These numbers are order-of-magnitude estimates due to the semi-empirical nature of matrix element scaling with the bond length and the neglected  ellipticity of atomic motion. To determine more precise values of these parameters, additional first-principles calculations are warranted.

For $\hbar\omega/\Delta\approx0.004$, orbital moment of the order of $10^{-2}$ $\mu_B$ estimated in Ref.~\cite{Basini2024} requires amplitude $Q\approx 0.08$~nm, which is unrealistically large despite STO's proximity to ferroelectric phase transition. Below, we discuss another, inter-atomic contribution that constructively combines with the discussed intra-atomic effect. Other possible contributions not captured by our MO model are also expected to constructively add, by angular momentum conservation argument. Furthermore, impurity states can be qualitatively interpreted in terms of a reduced effective gap $\Delta$ in Eq.~(\ref{eq:moment}), which can result in a significant contribution to transient magnetic moment. Altogether, these effects may account for the observed magnitude of induced magnetization.

\subsection{Transient interatomic orbital magnetism}

\begin{figure}
	\centering
	\includegraphics[width=0.9\columnwidth]{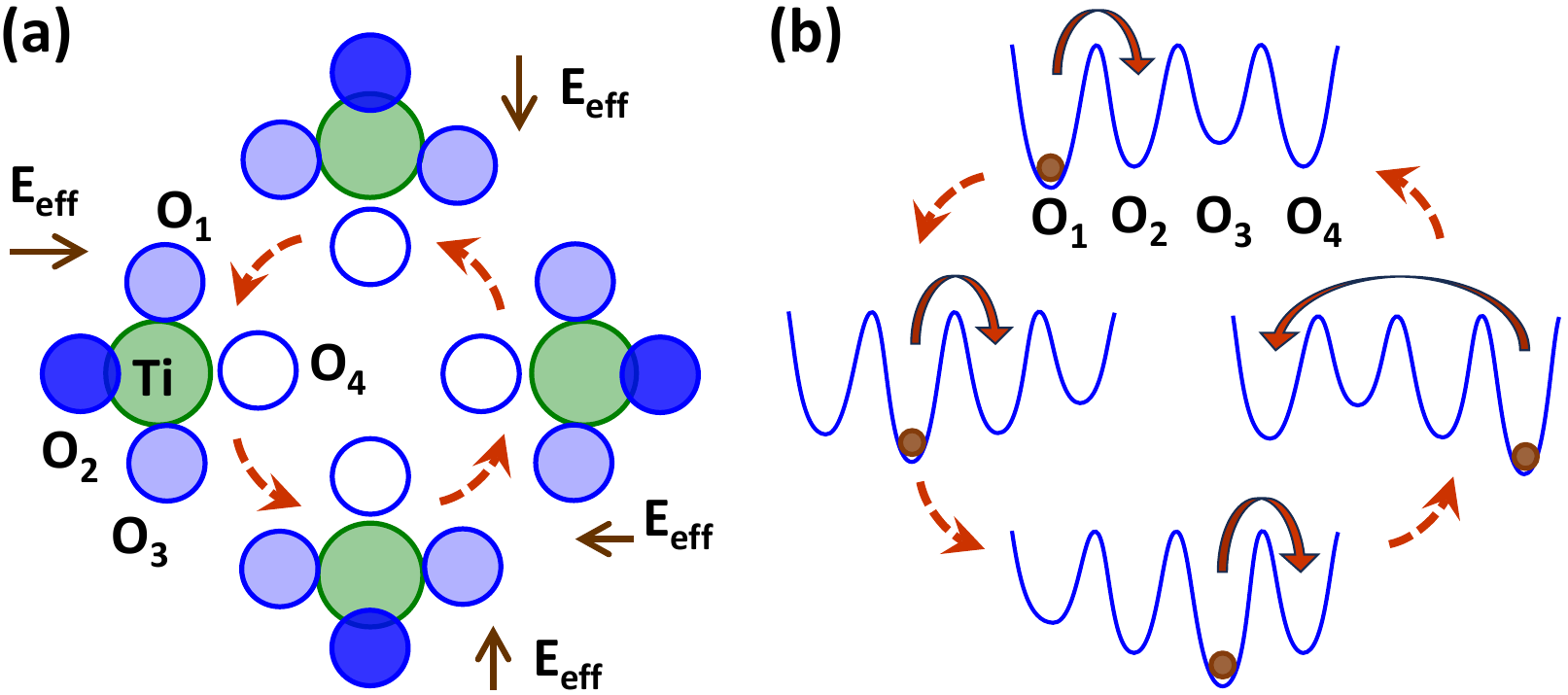}
	\caption{\label{fig:pumping} (a) Interatomic orbital magnetization in the MO approximation results from the charge re-distribution among the oxygen atoms coordinating Ti, due to the modulation of hybridization matrix elements. The color intensity represents electron density. (b) Analogy to the Thouless pump. The dots show the location of maximum electron density.}
\end{figure}

In the modern theory of orbital magnetism, atomic orbital magnetization discussed above generally co-exists with interactomic contribution~\cite{Resta2010}. Optical phonons break both time and spatial inversion symmetries, which is generally expected to result in a non-vanishing Berry curvature underlying this contribution~\cite{PhysRevB.100.054408}. Berry curvature of electronci bands is usually interpreted as a geometric effect in the reciprocal space. Nevertheless, one can ask whether the minimal real-space MO model introduced in this work can account for this mechanism, providing a simple estimate for its significance. Qualitatively, the MO wavefunction $\psi(\mathbf{Q})$ has a non-trivial orbital structure parametrized by the 2D manifold of ionic displacement field $\mathbf{Q}=(Q_x,Q_y)$, enabling a finite Berry curvature $\Omega_{xy}=2Im(\langle\partial\psi/\partial_{Q_y}|\partial\psi/\partial_{Q_x}\rangle)$~\cite{Zhang2023}.

The interatomic magnetic moment in the MO model results from the charge re-distribution due to the variation of Ti-O bonds. The closest to Ti oxygen atom is most strongly hybridized, resulting in the largest electronic density. In the top panel of Fig.~\ref{fig:pumping}(a), this is the top oxygen atom. Conversely, the bottom oxygen in this panel has the smallest charge. This can be interpreted as electron charge polarization due to the effective electric field $E_{eff}$ directed downward, opposite to the ionic polarization. As the displacement vector rotates, the effective field rotates together with it, resulting in a chiral motion of this polarization charge, i.e. a circulating current. 

The amount of charge transferred around the four oxygen atoms is independent of the period, consistent with the geometric origin of this process. It can be interpreted by analogy to the Thouless pump~\cite{PhysRevB.27.6083}, as illustrated in Fig.~\ref{fig:pumping}(b). The four-well profile represents the dependence of crystal potential on direction, with the minima along the Ti-O bonds labeled O$_1$-O$_4$ in Fig.~\ref{fig:pumping}(b). As the displacement vector rotates, this energy landscape is periodically modulated, resulting in charge pumping as indicated by a filled dot and curved arrows. In contrast to the Thoughless pump, finite size results in transfer of non-integer charge per cycle.

To analyze this effect, we use the time-dependent perturbation approximation for the MO states discussed above. We will show that the perturbed non-bonding states $\psi_{n1,n2}$ acquire a chiral amplitude on the oxygen atoms, resulting in a circulating current that produces interatomic magnetization. The lowest-order correction to the non-bonding wavefunction $\psi_{n1}$ is
\begin{equation}
\psi^{(1)}_{n1}=-\sum_{\sigma}V_{\sigma,n1}\frac{e^{i(\Delta/\hbar-\sigma\omega)}}{\Delta-\sigma\hbar\omega}\psi_{a\sigma}.
\end{equation}
The resulting time-dependent amplitude $c_n$ on $n^{th}$ oxygen atom is 
\begin{equation}
\begin{split}
c_n\approx\frac{1}{2}
-\frac{2\sin\theta_aQa_+e^{i\Delta t}}{\Delta^2}\\
*[(\Delta-\omega)e^{i(c\omega t-\pi n/2)}+(\Delta+\omega)e^{-i(c\omega t-\pi n/2)}].
\end{split}
\end{equation}
According to this expression, the wavefunction acquires a chiral phase winding around the oxygen atoms, resulting in a circulating current between sites $n$ and $n+1$
\begin{equation}
I_{n,n+1}=-\frac{2et_{O-O}}{\hbar}Im(c^*_nc_{n+1}),
\end{equation}
where $e$ is the magnitude of electron charge. This current includes an oscillating component and a constant contribution
\begin{equation}
\langle I_{n,n+1}\rangle=-\frac{32et_{O-O}\sin^2\theta_aQ^2a^2_+\omega c}{\Delta^3}.
\end{equation}
Since the current is proportional to $\omega$, the charge transferred per period is independent of $\omega$, consistent with the geometric origin of this effect. The magnetic moment is
\begin{equation}
m_{z1}=-\frac{a^2}{4}\langle I_{n,n+1}\rangle=\frac{8ea^2t_{O-O}\sin^2\theta_aQ^2a^2_+\omega c}{\Delta^3}
\end{equation}
where $a=0.39$~nm is the lattice constant of STO. Similar analysis for $\psi_{n2}$ yields
\begin{equation}
m_{z2}=\frac{8ea^2t_{O-O}\sin^2\theta_aQ^2a^2_-\omega c}{\Delta^3}.
\end{equation}
Since $a_+>a_-$, the two contributions do not cancel, resulting in a non-vanishing total orbital moment. Taking into account the two spin directions, we obtain the total inter-atomic orbital moment
\begin{equation}\label{eq:mzia}
\mu_{z,ia}=\frac{32\mu_1c\hbar\omega\sin^2\theta_aQ^2_0(a^2_+-a^2_-)}{\Delta^3}.
\end{equation}
the same expression as Eq.~(\ref{eq:moment}) for the atomic moment, except it is scaled by the magnetic moment $\mu_1=-ea^2t_{O-O}/2\hbar$ instead of the Bohr magneton.
In particular, the induced moment is quadratic in displacement and is transient, consistent with the experimental observations. We use Koster-Slater approximation to estimate $t_{O-O}=-0.8$~eV, giving $\mu_1\approx1.6\mu_B$. It is remarkable that two contributions of different origin and expressed in terms of unrelated material parameters give such similar results. This is consistent with other studies suggesting that intra- and inter-atomic contributions to orbital magnetization are usually comparable~\cite{PhysRevB.106.104414}.

\section{Discussion and Summary} 

We showed that a minimal molecular orbital approximation can serve as a testbed for the mechanisms contributing to magnetism induced by interaction between lattice dynamics and electrons, enabling simple estimates of the relevant contributions. Chiral lattice dynamics produces intra-atomic orbital electron magnetization due to the polarization of the Ti atom, as well as inter-atomic contribution due to the chiral current circulating around oxygen atoms coordinating it. The magnetization estimated in our model is smaller than experimentally observed. Remarkably, the inter-atomic contribution is very close to intra-atomic contribution. This suggests that additional contributions not captured by our model may also be similar. Such contributions may include the current circulating around the square plaquettes whose vortexes are formed by the Ti atoms, as well as orbitals $d_{xy}$, $p_x$ and $p_z$ and the corresponding bonds. In contrast to ionic motion, different electronic contributions should add constructively based on the general angular momentum conservation argument, which altogether may account for the observed large moment.

Our model of STO provides a tractable example of non-perturbative electron-phonon interaction that cannot be interpreted in the language of electron-phonon scattering, supporting the proposed strong-coupling theories~\cite{Mentink2019}. The transient nature of induced electronic magnetism shows that the demonstrated effects describe dressing of chiral phonons by electron dynamics, resulting in re-normalization of the phonon magnetic moment. A similar conclusion was reached in the analysis of geometric phonon magnetism in gapped bilayer graphene, where non-geometric phonon magnetic moment is absent~\cite{Zhang2023}. In contrast to the inverse cubic dependence on the gap obtained in out analysis, the magnetization in graphene was shown to scale inversely with the square of the gap, warranting further analysis and experimental studies. Our analysis also predicts that induced magnetization is quadratic in ionic displacement, instead of the linear dependence in Ref.~\cite{Zhang2023}, which is consistent with the experiment.

The demonstrated effect is a general consequence of cogwheel-like locking between the phase of atomic orbitals with finite orbital moments and the positions of coordinating atoms. Just as the static crystal field generally quenches orbital moments, dynamical crystal field results in modulation of their unquenched components, i.e. transient orbital moments. All the bands remain filled or empty in this process. Therefore, it is protected by the large gap $\Delta=3.2$~eV of STO and is robust with respect to thermal relaxation. Phonon-induced electron magnetization only requires a Kramer's doublet formed by orbitally nontrivial states, which is general to nonmagnetic materials with $p$ or $d$ bonding. However, its efficiency can be enhanced by degenerate orbital states whose superpositions can support large unquenched orbital moments. In our model, both the intra- and the inter-atomic moments, Eqs.~(\ref{eq:moment}), (\ref{eq:mzia}) are proportional to $(a^2_+-a^2_-)=a_la_t$. The coefficient $a_t$ reflecting orbital chirality would vanish for an orbitally quenched state, resulting in vanishing electronic orbital moment. In STO, a finite $a_t$ results from the orbital polarization of the antibonding states $d_+$, $d_-$. The condition for orbital degeneracy is naturally satisfied by the high-symmetry (e.g. cubic or tetragonal) phases of complex oxides. In systems with $C_3$ symmetry such graphene, this degeneracy is a direct consequence of the chirality of its irreducible representations.

In our analysis, the magnetic polarization scales with the gap as $1/\Delta^3$ at $\Delta\gg\hbar\omega$, resulting in a dramatic increase of efficiency at small gap. At $\Delta=\omega$, the orbital moment in the approximation Eq.~(\ref{eq:moment}) diverges, which corresponds to resonant excitation of the chiral state $\psi_{a,\sigma}$ with $\sigma=c$. As evidenced from Eq.~(\ref{eq:p_sigma_inf}), for $\hbar\omega$ approaching $\Delta$, the non-transient orbital moment also increases, signifying an increased population of orbitally polarized electrons excited into the conduction band. This dependence allows control of the induced magnetic moment by gap modulation~\cite{Zhang2023}. A similar effect can be achieved by doping~\cite{PhysRevB.107.L020406}. Efficient resonant enhancement was observed in narrow-gap materials~\cite{Luo2023}, and supported by analysis in terms of phonon-induced splitting of Kramers doublet~\cite{PhysRevB.110.094401,doi:10.1021/acsnano.4c18906}. This mechanism of controlled generation of orbital moments can be attractive for orbitronic applications, bypassing the electrical currents needed in the the orbital moment generation by orbital Hall effect~\cite{PhysRevLett.102.016601,Mokrousov2020}. Importantly, orbital magnetization can be transferred across interfaces between  different materials due to the orbital selectivity of hybridization~\cite{PhysRevB.108.L180404}, and used to control the magnetic state via SOC~\cite{PhysRevLett.125.177201,Lee2021}.

\textit{Acknowledgments.} This work was supported by the NSF award ECCS-2005786. I thank Stefano Bonetti and Alexander Balatsky for the helpful discussions, and Dominik Juraschek for the feedback on the manuscript.

\bibliography{STO_orb}
\bibliographystyle{apsrev4-2}
\end{document}